\documentclass[aps,pre,twocolumn,superscriptaddress,floatfix]{revtex4}
\usepackage{graphicx}
\usepackage[dvips,unicode,colorlinks,linkcolor=blue,citecolor=blue,urlcolor=blue]{hyperref}
\usepackage[english]{babel}
\usepackage{dcolumn}% Align table columns on decimal point
\usepackage{bm}% bold math
\usepackage{color}
\usepackage{amsmath}

\begin{document}
%\preprint{APS/123-QED}
\title{Temperature-induced reversal effects of kink dynamics in carbon nanotube on flat substrate}
\author{Alexander V. Savin}
\affiliation{ N.N. Semenov Federal Research Center for Chemical Physics,
Russian Academy of Science (FRCCP RAS), Moscow, 119991, Russia}
\affiliation{Plekhanov Russian University of Economics, Moscow, 117997 Russia}
%\email{asavin@center.chph.ras.ru}
%\author{\color{red}Leonid I. Manevitch}
%\affiliation{ N.N. Semenov Federal Research Center for Chemical Physics,
%Russian Academy of Science (FRCCP RAS), Moscow, 119991, Russia}
\author{Margarita Kovaleva}
\affiliation{ N.N. Semenov Federal Research Center for Chemical Physics,
Russian Academy of Science (FRCCP RAS), Moscow, 119991, Russia}
%\email[]{margo.kovaleva@gmail.com}
\affiliation{HSE University, Moscow, 109028 Russia}

\begin{abstract}
Carbon nanotubes are nano-objects with quite anisotropic properties, for example the mechanical 
properties in longitudinal and radial directions differ significantly. This feature of the carbon 
nanotubes yields many interesting phenomena investigated in last decades. One of them is 
the ability to form both hollow and collapsed states if the radius of the nanotube is 
large enough. The transitions between the two states have been also reported. 
In our study we present single-walled carbon nanotube interacting with a plane substrate 
and characterize the energy of interaction with the substrate using effective Lennard-Jones-type potential.
We show energy of the homogeneous open and collapsed states depending on the radius of the 
carbon nanotube and report on the bi-stability in some range of the nanotube diameters.  
Using the molecular-dynamical simulations we look at the evolution of the initial half-opened, 
half-collapsed state and demonstrate that the transition area from one state to another 
is spatially localized having features of topological soliton (kink or anti-kink).
We show that the value and the direction of the kink propagation speed depend
significantly on the nanotube diameter as well as on the temperature of the system. 
We also discuss the mechanism of the process using a simplified model with asymmetric double-well 
potential and show the entropic nature of the transition.
\end{abstract}
\pacs{44.10.+i, 05.45.-a, 05.60.-k, 05.70.Ln}
\maketitle

\section{Introduction}\label{Int}

Carbon nanotubes (CNTs) are cylindrical macromolecules with a diameter varying from half a 
nanometer up to 20 nanometers. They are long, hollow tubule structures made of graphene sheets.
Similar structures were obtained firstly more than 70 years ago during the thermal decomposition
of carbon monoxide on an iron contact \cite{Radushkevich52}. However, CNTs themselves were
synthesized only 30 years ago as by-products of fullerene C$_{60}$ synthesis \cite{Iijima91}.
CNTs are promising engineering nano-materials with increasing usage and significance in nanotechnology
\cite{Eleckii02}. CNTs  attract interest of researchers in physics, material science, electronics
and biotechnology and nanotechnology due to their unique thermal, mechanical, optical and
biological properties \cite{Ferreira16,Menezes19,Venkataraman19}.
From the mechanical point of view nanotube is a quasi-one-dimensional molecular structure with
pronounced nonlinear properties \cite{Dresselhaus01,Astakhova01,Savin04,Savin08}.

Nanotubes have high longitudinal (axial) and relatively weak transverse (radial) stiffness.
Because of this, at sufficiently large diameters, nanotubes due to the weak non-covalent 
interaction of atoms can transform from a hollow cylindrical shape to a collapsed state 
\cite{Chopra95,Gao98,Xiao07,Baimova15,Xu2019,Grande2020,Magnin2021,Umeno}. Non-valent interaction with the substrate can also change the cylindrical shape of the nanotube \cite{Hertel98,Xie10,Yuan18}.
It has been shown that long multi-walled narrow graphene nanoribbons can be created by squashing
carbon nanotubes using a thermally assisted high-pressure process \cite{Chen21,Toh21}.
Such collapsed nanotubes can be used as semiconducting graphene nanoribbons.

We report on the collapse of the carbon nanotube on the substrate, and underline that this process 
is controlled by the temperature of the system. Moreover, we show that the effect is reversal, 
which means that the nanotube that collapsed due to temperature decrease can open back 
if the temperature is increased again. The front of the opening or collapsing in the 
semi-collapsed nanotube has the soliton-like profile, which can diffusively move in the 
longitudinal direction. Using a simplified (biparabolic) effective model 
for the evolution of the front, we show that Brownian motion of a kink in asymmetric 
double-well potential can describe the process. 
Using this approach we compare the theoretically obtained speed 
of the solitary wave with the speed of the front from the molecular-dynamical investigation 
and obtain a fairly good agreement in the low-temperature range.

\section{Model}\label{Model}

We consider a CNT with chirality indices $(m, m)$. The cylindrical structure of
such a nanotube is formed by periodic repetition along the $x$ axis of transverse cyclic zigzag
chains of $K=4m$ carbon atoms:
\begin{eqnarray}
x_{n,(j-1)m+i}&=&h_i+a(n-1),\nonumber\\
y_{n,(j-1)m+i}&=&R\cos(\phi_i+(j-1)\Delta\phi),\nonumber\\
z_{n,(j-1)m+i}&=&R\sin(\phi_i+(j-1)\Delta\phi),\nonumber
\end{eqnarray}
where the first index $n=0,\pm1,\pm2,...$ numbers the transverse rings of atoms (unit cells),
the second index $k=(j-1)m+i$, $j=1,...,m$, $i=1,2,3,4$, -- atoms in these rings.
Here the angular pitch $\Delta\phi=2\pi/m$, the radius of the nanotube $R=r_0/2\sin(\Delta\phi/6)$
($r_0=1.418$\AA~ is the equilibrium length of the C--C valence bond in a graphene sheet),
the longitudinal pitch $a=r_0\sqrt{3}$, the longitudinal displacements $h_1=h_4=0$, $h_2=h_3=a/2$,
the angular displacements $\phi_1=0$, $\phi_2=\Delta\phi/6$, $\phi_3=3\Delta\phi/6$, $\phi_4=4\Delta\phi/6$.

Let the index $n$ number the transverse ring elements of the CNT with the armchair structure,
then the positions of the atoms of each ring can be described by a vector of $3K$ dimension
${\bf u}_n=\{ {\bf v}_{n,k}=(x_{n,k},y_{n,k},z_{n,k})\}_{k=1}^K$.
The Hamiltonian of a nanotube has the following form
%-------------------------------------- f1--------------------------------
\begin{equation}
H=\sum_{n=-\infty}^{+\infty}\sum_{k=1}^K\left\{\frac12M(\dot{\bf v}_{n,k},\dot{\bf v}_{n,k})+P_{n,k}+V(z_{n,k})\right\},
\label{f1}
\end{equation}
%-------------------------------------- f1--------------------------------
where $M$ is the mass of a carbon atom, ${\bf v}_\alpha=\left(x_\alpha(t),y_\alpha(t),z_\alpha(t)\right)$
is a vector defining the position of the atom with two-dimensional index $\alpha=(n,k)$ at time $t$.
Term $P_\alpha$ describes the interaction of the atom with the index $\alpha$ with the other atoms
of the nanotube, the last term of $V(z_\alpha)$ sets the energy of interaction of the atom
with flat substrate.
%----------------------------------------------------------------
\begin{figure}[tb]
\includegraphics[angle=0, width=1.0\linewidth]{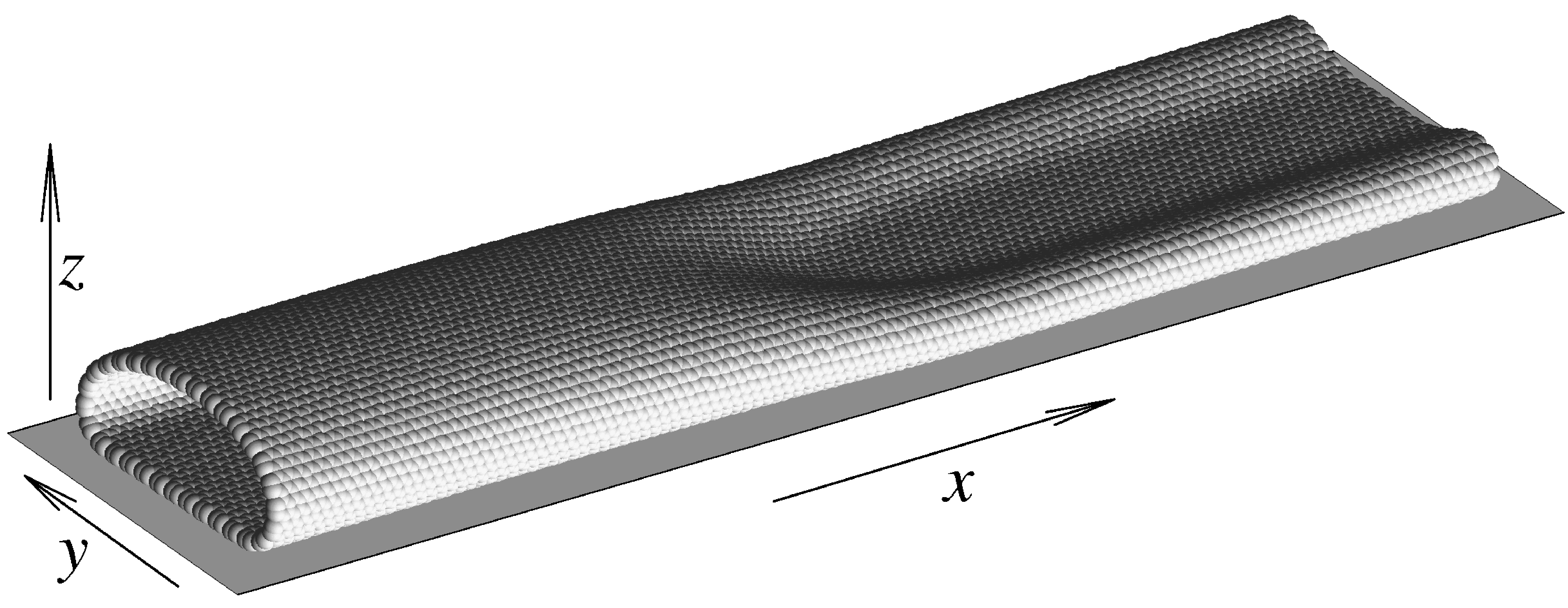}
\caption{\label{fig1}\protect
View of a carbon nanotube with a chirality index (31,31) located on a flat substrate formed
by the surface of a silicon carbide crystal 6HSiC (0001). The left end of the nanotube is
in an open stationary state (the cross-section of the nanotube has the shape of a convex drop),
the right end is in a collapsed stationary state (the cross-section has the shape of an asymmetric
dumbbell with a flat two-layer central part).
In the middle part of the nanotube, a localized
region of its smooth transition from one stable stationary state to another is formed.
}
\end{figure}
%----------------------------------------------------------------

To describe the carbon-carbon valence interactions we use a standard set of
molecular dynamics potentials \cite{Savin10}. The valence bond between two neighboring carbon
atoms $\alpha$ and $\beta$ can be described by the Morse potential
%-------------------------------------- f2--------------------------------
\begin{equation}
U_1({\bf v}_\alpha,{\bf v}_\beta)=\epsilon_1\{\exp[-\alpha_0(r_{\alpha\beta}-r_0)]-1\}^2,
\label{f2}
\end{equation}
%-------------------------------------- f2--------------------------------
where $r_{\alpha\beta}=|{\bf v}_\alpha-{\bf v}_\beta|$, 
$\epsilon_1=4.9632$~eV is the valence bond energy and $r_0=1.418$~\AA~ is the equilibrium valence bond
length.

Valence angle deformation energy between three adjacent carbon
atoms $\alpha$, $\beta$, and $\gamma$ can be described by the potential
%-------------------------------------- f3--------------------------------
\begin{equation}
U_2({\bf v}_\alpha,{\bf v}_\beta,{\bf v}_\gamma)=\epsilon_2(\cos\varphi-\cos\varphi_0)^2,
\label{f3}
\end{equation}
%-------------------------------------- f3--------------------------------
where
$\cos\varphi=({\bf v}_\alpha-{\bf v}_\beta,{\bf v}_\gamma-{\bf v}_\beta)/r_{\alpha\beta}r_{\gamma\beta}$,
and $\varphi_0=2\pi/3$ is the equilibrium valent angle.
Parameters $\alpha_0=1.7889$~\AA$^{-1}$ and $\epsilon_2=1.3143$~eV can be found from the small amplitude
oscillations spectrum of the graphene sheet \cite{Savin08}.

Valence bonds between four adjacent carbon atoms $\alpha$, $\beta$, $\gamma$, and $\delta$
constitute the torsion angles, the potential energy of which can be defined as
%-------------------------------------- f4--------------------------------
\begin{equation}
U_3(\phi)=\epsilon_3(1-\cos\phi),
\label{f4}
\end{equation}
%-------------------------------------- f4--------------------------------
where $\phi$ is the corresponding torsion angle ($\phi_0=0$ is the equilibrium value of the angle)
and $\epsilon_3=0.499$~eV \cite{Gunlycke08}.

More detailed discussion and motivation of our choice of the interaction potentials (\ref{f2}),
(\ref{f3}), (\ref{f4}) can be found in our earlier publication \cite{Savin10}.

Non-valent interactions of carbons atoms are described \cite{Setton96} by the Lennard-Jones
potential
%-------------------------------------- f5--------------------------------
\begin{equation}
V_{cc}(r)=4\epsilon_{cc}[(\sigma_{cc}/r)^{12}-(\sigma_{cc}/r)^{6}],
\label{f5}
\end{equation}
%-------------------------------------- f5--------------------------------
where $\epsilon_{cc}=0.002757$~eV is the binding energy and $\sigma_{cc}=3.393$~\AA.
The potential attains a minimum value of $-\epsilon_{cc}$ at $r_{cc}=2^{1/6}\sigma_{cc}=3.807$~\AA~
(equilibrium interatomic distance).

To simulate the dynamics of a nanotube located on a flat substrate formed by the surface of
a molecular crystal, it is necessary to find the interaction potential of the carbon atom with
the substrate. For this purpose, the interaction energy of a finite flat sheet of graphene with
a flat surface of a 6H-SiC(0001) silicon carbide crystal was found \cite{Sforzini15}.
The calculations used a graphene sheet size $1.985\times1.643$~nm$^2$, consisting of 160 carbon
atoms, located parallel to the surface of the crystal $z=0$ at a distance of $h$.
The interaction energy of each carbon atom with a silicon carbide crystal was calculated
as the sum of the Lennard-Jones potentials (\ref{f5}) with the values of the parameters from \cite{Rappe92}.
At each value of $h$, the energy of the interaction of the sheet with the crystal was averaged
along the shift along the $x$ and $y$ axes, and then normalized by the number of atoms
in the sheet. As a result, the dependence of the interaction energy of one atom of the sheet
on its distance to the substrate plane $V(h)$ was obtained.

The calculations showed that the interaction energy with the substrate $V(h)$ can be described
with high accuracy by the $(k,l)$ Lennard-Jones potential
%-------------------------------------- f6--------------------------------
\begin{equation}
V(h)=\epsilon_0[k(h_0/h)^l-l(h_0/h)^k]/(l-k),
\label{f6}
\end{equation}
%-------------------------------------- f6--------------------------------
where degree $k=3.75$, $l=17$, binding energy $\epsilon_0=0.073$~eV, equilibrium distance
to plane $h_0=4.19$~\AA.

\section{Homogeneous stationary states of the nanotube}\label{Homstat}

To find a uniform stationary state of the nanotube it is necessary to solve the problem
of the minimum potential energy
%-------------------------------------- f7--------------------------------
\begin{equation}
E=\sum_{k=1}^K \{P_{(0,k)}+V(z_{(0,k)})\}\rightarrow\min:\{{\bf v}_{0,k}\}_{k=1}^K,~a
\label{f7}
\end{equation}
%-------------------------------------- f7--------------------------------
by the coordinates of atoms in one unit cell $(n=0)$ and the value of the longitudinal period.
Problem (\ref{f7}) was solved numerically by the conjugate gradient method. By choosing the
initial configuration of atoms in the cell, all stable stationary states of the nanotube can be obtained.

The solution of the problem (\ref{f7}) showed that at the index $m<17$ (at the radius of the
nanotube $R<11.517$~\AA), the nanotube $(m,m)$ on a flat substrate has only one stable stationary
state (open ground state), in which the cross-section of the nanotube has the form of a convex
drop -- see Fig.~\ref{fig1}.
At $m\ge 17$, in addition to the open state, there is also a second stable
stationary state (collapsed ground state), in which the cross-section has the form of
an asymmetric dumbbell with a flat two-layer central part.
%----------------------------------------------------------------
\begin{figure}[tb]
\begin{center}
\includegraphics[angle=0, width=1\linewidth]{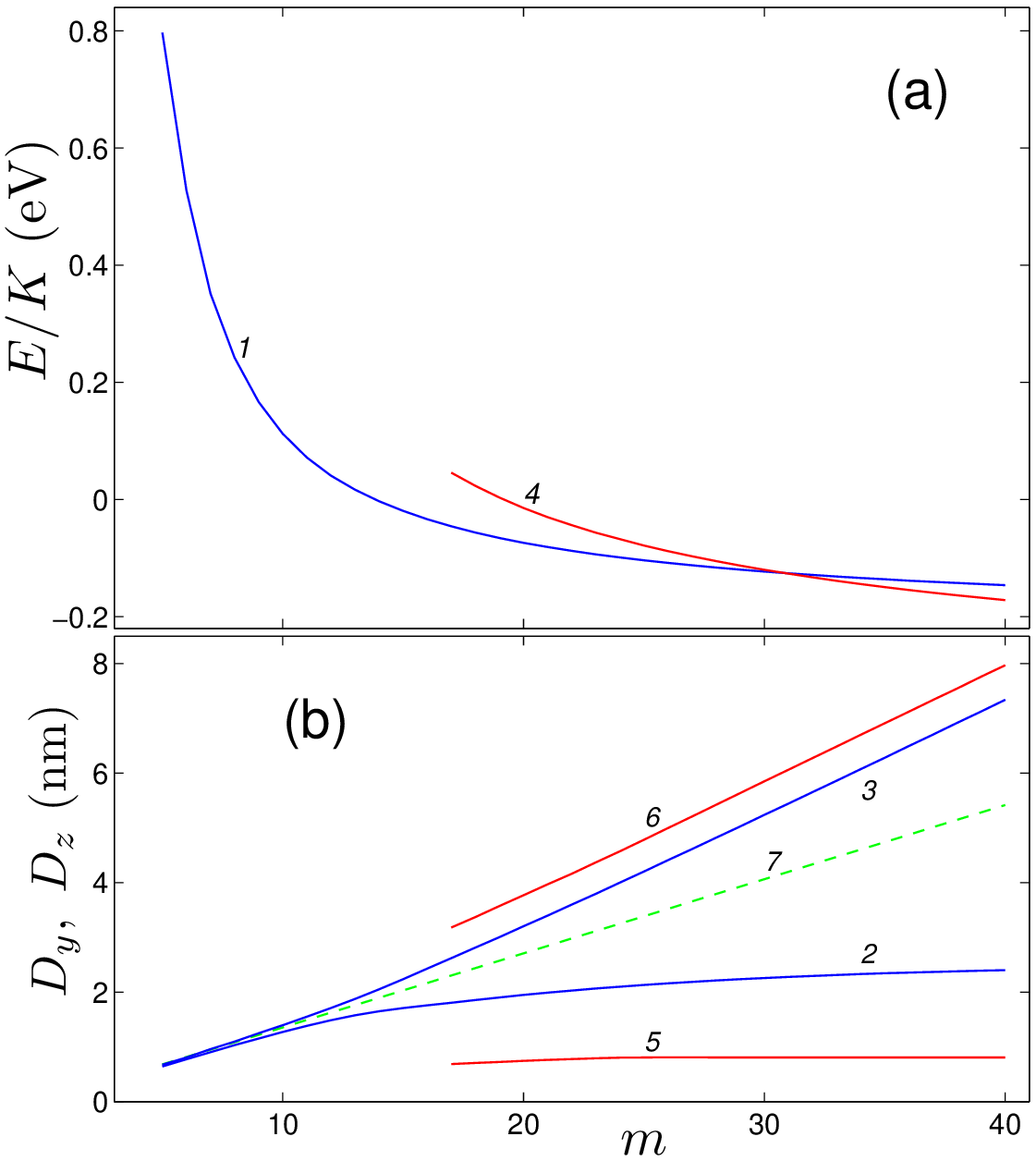}
\end{center}
\caption{\label{fig2}\protect
Dependence of (a) normalized energy $E/K$ and (b) diameters $D_y$, $D_z$ nanotubes with chirality
index $(m,m)$ from the value of the index $m$. Curves 1, 2, 3 give the dependence for an open
primary state, and curves 4, 5, 6 -- the represent the ground state of the collapsed nanotube
located on a flat substrate of silicon carbide 6H-SiC(0001). Line 7 gives the diameter dependence
for an isolated cylindrical nanotube.}
\end{figure}
%----------------------------------------------------------------

The stationary state of the nanotube $(\{{\bf v}_{0,k}^0\}_{k=1}^K,a)$ is characterized
by its normalized energy $E/K$, transversal and vertical diameters
$$
D_y=\max_{k_1,k_2}|y_{0,k_1}^0-y_{0,k_2}^0|,~~
D_z=\max_{k_1,k_2}|z_{0,k_1}^0-z_{0,k_2}^0|.
$$

The dependencies of the normalized energy $E/K$ and diameters $D_y$, $D_z$ on the value of the
index $m$ of the nanotube located on a flat substrate of crystalline silicon carbide are presented
in Fig.~\ref{fig2}. As can be seen from the figure, on the substrate the vertical diameter of the
nanotube is always smaller than the transverse diameter ($D_z<D_y$) -- interaction with the
substrate leads to flattening of the nanotube. The transverse diameter of the stationary state
always grows in proportion to the index value: $D_y\sim m$ for $m\rightarrow\infty$.
The vertical diameter of the open
state of the nanotube grows as the logarithm of the index: $D_z\sim \ln m$, and the vertical
diameter of the collapsed state remains constant $D_z\approx 8.07$~\AA~ (the diameter value
is determined by the height of the end curves of the cross-section of the nanotube).

At low values of the index $17\le m\le 30$, the open state is more advantageous in energy than
the collapsed state of the nanotube. If $m\ge 31$ the ground state becomes the collapsed one.
With an increase in the index $m$, the difference in the energy of the states monotonically
decreases at $m<31$, reaches a minimum value  $\Delta E=|E_o-E_c|=0.014$~eV at $m=31$
(open state energy value is $E_o=-3.909$~eV and collapsed state energy value is $E_c=-3.923$~eV)
and monotonically increases at $m>31$ [see Fig.~\ref{fig2}~(a)].
%----------------------------------------------------------------
\begin{figure}[tb]
\begin{center}
\includegraphics[angle=0, width=1\linewidth]{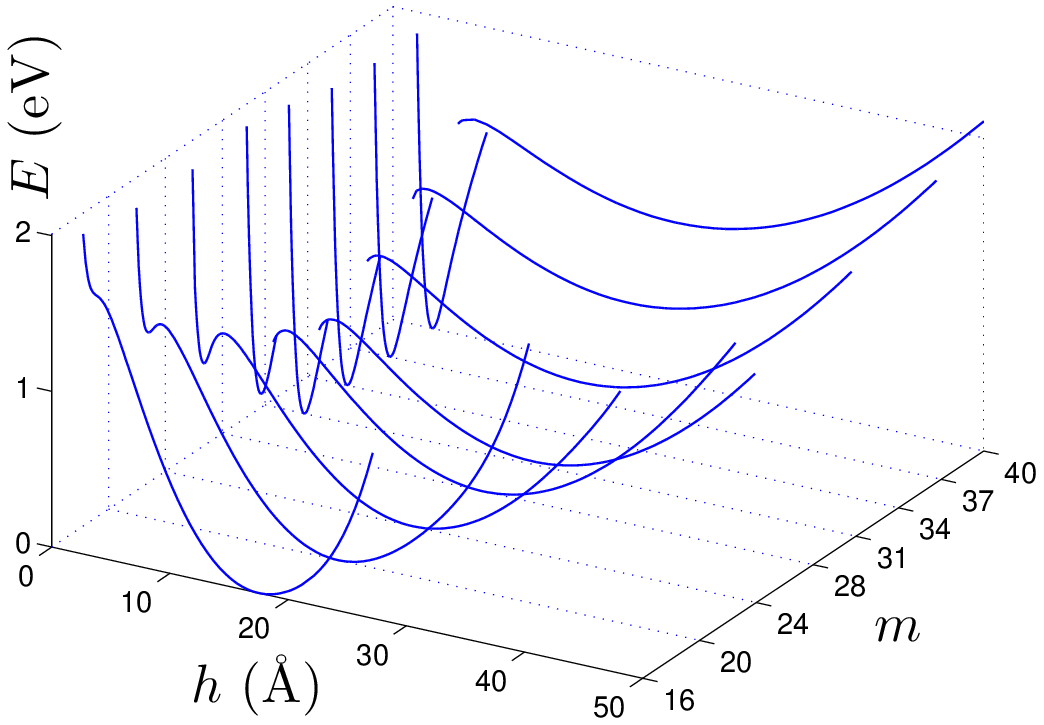}
\end{center}
\caption{\label{fig3}\protect
Dependence of the energy $E$ of the nanotube $(m,m)$ located on the substrate from the flat
surface of the silicon carbide crystal 6H-SiC (0001) on the distance $h$ from the substrate
to the center of the opposite side of the nanotube at different values of the index $m$.
The energy is counted from the minimum value.
}
\end{figure}
%----------------------------------------------------------------

\section{Energy profile of the transition between the stationary states}\label{Enprof}

The results obtained allow us to conclude that a single-walled nanotube with an index $m\ge 16$
is a bistable system with two stable (open and collapsed) states.
The energy profile of the transition between these two states can be found numerically;
for that we
solve the problem on the minimum of (\ref{f7}) for each fixed value of the distance $h$
between the substrate plane and the center of the upper side of the nanotube.
The numerical solution of this problem allows us to obtain the dependence $E(h)$, describing
the change in the energy of the nanotube during its homogeneous collapse.

The view of potential $E(h)$ at different values of the index of the nanotube $m$
is shown in Fig.~\ref{fig3}.
As can be seen from the figure at $m<17$, the function $E(h)$ is a one-well potential
with a minimum corresponding to the open state. At $m=17$, the potential obtains a two-well form,
it has a new narrow minimum corresponding to the collapsed stationary state.
With the increase of the index $m$ (with the increase of the radius) of the nanotube, the depth
of the new narrow minimum monotonically increases.
At $m\ge 31$ this minimum becomes energetically preferred,
i.e. the collapsed state becomes the main state of the CNT.
In our case the energy profile $E(h)$ has the form of a strongly asymmetric
double-well potential with a narrow first valley and a wide second valley.

The transition of a nanotube from one stationary state to another can be described qualitatively
as the motion of a kink (topological soliton) in the $\phi$--4 model with an asymmetric
double-well potential $E(h)$ having one narrow deep well corresponding to the collapsed state
and a second wide well with higher energy corresponding to the open state of the nanotube.
This case is considered in detail in \cite{Savin97,Costantini01}, where it is shown that
the direction of motion of the kink in such a chain depends on the temperature value.
The kink motion describes the sequential transition of a chain from a non-ground state
to a ground state.
At low temperatures, the main state will always be in a deeper narrow valley, at high temperatures
-- a higher energy minimum but in wider second valley. Switching the ground state with an increase in
temperature leads to a change in the direction of motion of the topological soliton.
Let's model this effect for motion along the nanotube topological soliton describing
its sequential transition from one ground state to another -- see Fig.~\ref{fig1}.
For an isolated nanotube, the effect of changing the direction of motion of this topological
soliton with temperature changes was detected in \cite{Chang10}, but did not receive sufficient explanation.
%----------------------------------------------------------------
\begin{figure}[t]
\begin{center}
\includegraphics[angle=0, width=1\linewidth]{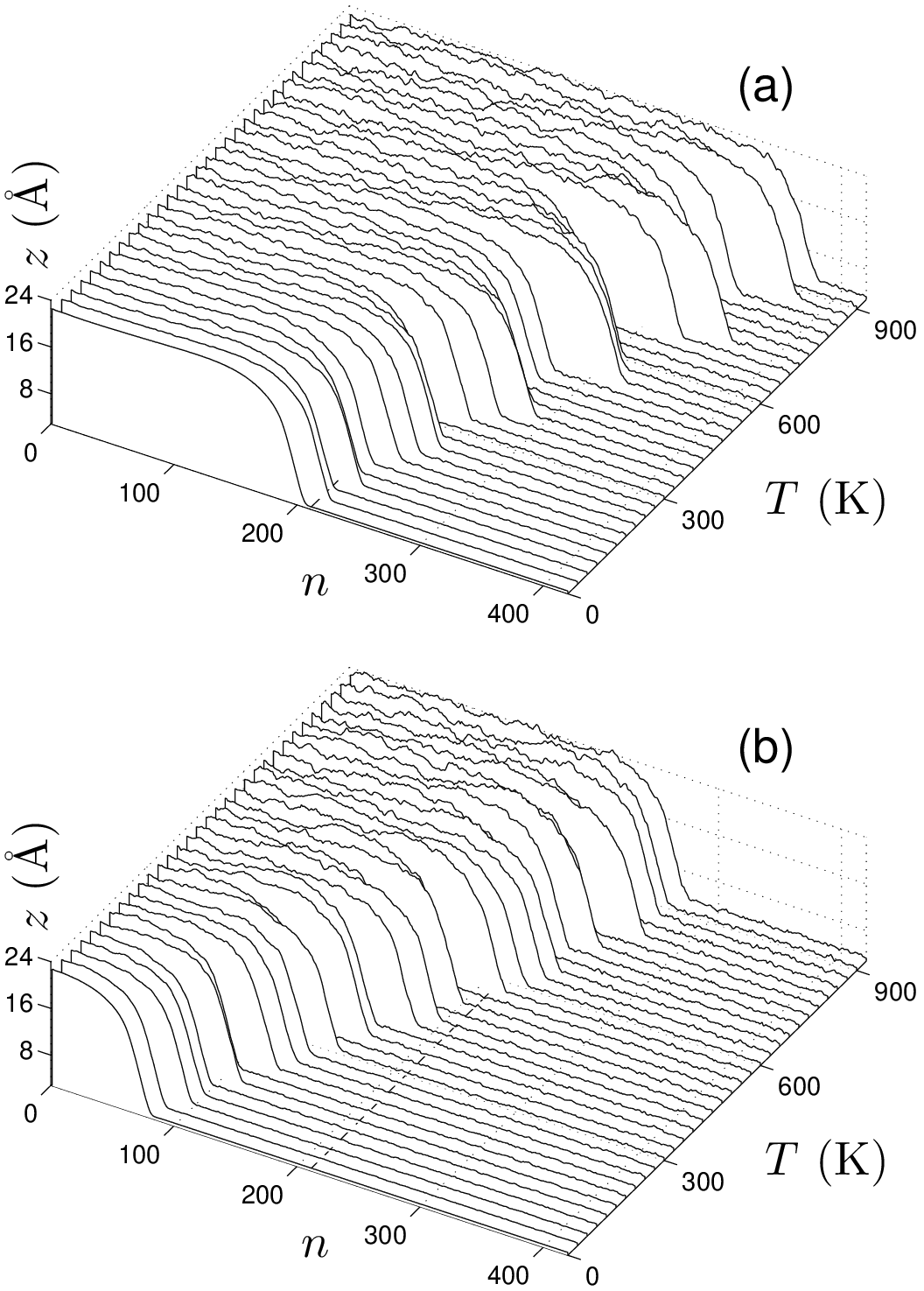}
\end{center}
\caption{\label{fig4}\protect
View of the upper part of the longitudinal section of the nanotube $(m,m)$ with a topological
soliton at index  (a) $m=31$ and (b) $m=32$. The view of the section is shown at thermostat
temperatures $T=0,30,...,930$~K at time $t=400$ and $t=500$~ps. the dotted line shows the
position of the center of the soliton at the initial time value.
}
\end{figure}
%----------------------------------------------------------------
\section{Temperature dependence of the direction of motion of a topological soliton}
\label{TempDep}

%----------------------------------------------------------------
\begin{figure}[t]
\begin{center}
\includegraphics[angle=0, width=1\linewidth]{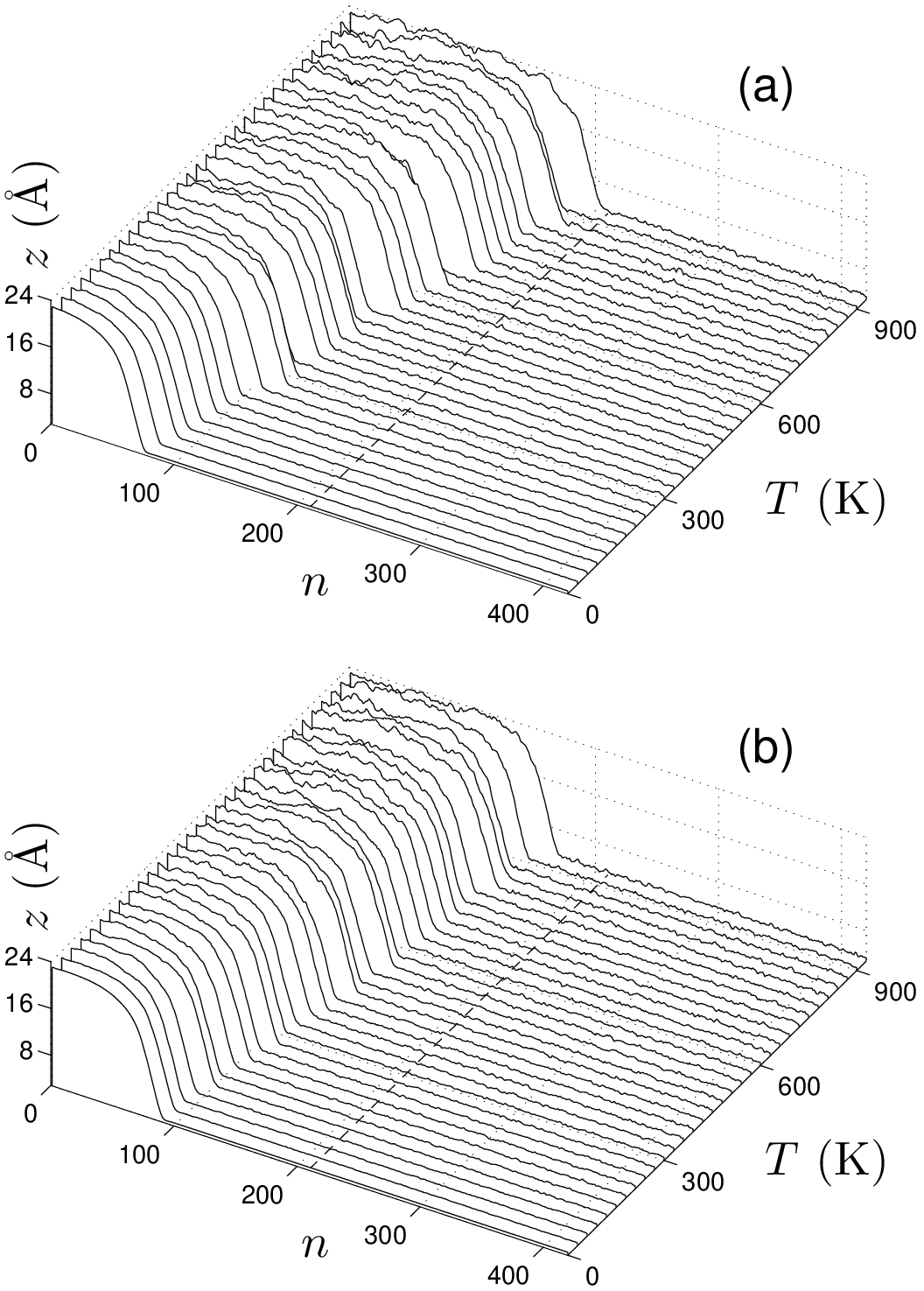}
\end{center}
\caption{\label{fig5}\protect
View of the upper part of the longitudinal section of the nanotube $(m,m)$ with a topological
soliton at index  (a) $m=33$ and (b) $m=34$. The view of the section is shown at thermostat
temperatures $T=0,30,...,930$~K at time $t=300$ and $t=200$~ps. The dotted line shows
the position of the center of the soliton at the initial time value.
}
\end{figure}
%----------------------------------------------------------------
To model the motion of a topological soliton in a thermalized nanotube, we consider a finite
nanotube with a chirality index $(m,m)$ of $N=420$ cross elements. At the initial moment of time,
the left part of the nanotube $1\le n\le 180$ will be transferred to the hollow stationary state,
the right part $N-180<n\le N$ -- to the collapsed state, and in the central part $180<n\le N-180$
we will define a linear continuous transition from one state to another. We fix the position
of atoms in the end cells of the nanotube $n=1,2$ and $n=N-1,N$ and consider the dynamics of
the nanotube immersed in the Langevin thermostat.

The dynamics of a thermalized nanotube is described by a system of Langevin equations
%-------------------------------------- f8--------------------------------
\begin{eqnarray}
M\ddot{\bf v}_{n,k}=-\frac{\partial H}{\partial {\bf v}_{n,k}}-\Gamma M\dot{\bf v}_{n,k}+\Xi_{n,k},
\label{f8}\\
n=3,4,...,N-2,~~k=1,2,...,K, \nonumber
\end{eqnarray}
%-------------------------------------- f8--------------------------------
where the vector ${\bf v}_{n,k}=(x_{n,k},y_{n,k},z_{n,k})$ specifies the coordinates of the $k$-th
atom in the $n$-th transversal element,
$H$ is the Hamiltonian (\ref{f1}), $\Gamma=1/t_r$ is the friction
coefficient (velocity relaxation time is $t_r=1$~ps), and $\Xi_{n,k}=(\xi_{n,k,1},\xi_{n,k,2},\xi_{n,k,3})$
is three-dimensional vector of the normally distributed random forces,
normalized by the conditions
$$
\langle\xi_{n_1k_1j_1}\!(t_1)\xi_{n_2k_2j_2}\!(t_2)\rangle\! =
\!2M\Gamma k_BT\delta_{n_1n_2}\delta_{k_1k_2}\delta_{j_1j_2}\delta(t_1-t_2)
$$
($k_B$ is the Boltzmann constant, $T$ is thermostat temperature).

If thermostat temperature is $T=0$, than initially given topological defect in nanotube,
with the center at the node $N/2$ will always move to the right, translating the main part
of the nanotube in the  hollow state,
when the index $m\le30$. Same way, the defect will move to the left,
shifting the nanotube to close state if the condition $m>30$ holds.
In this case, a localized region is formed in the nanotube in which there is a smooth transition
from one stationary state to another -- see Fig.~\ref{fig1}. This region moves at a constant speed
while maintaining its shape, i.e. behaves like a topological soliton. The upper part of the
central longitudinal section of the nanotube has the shape of an anti-kink $\{z_n\}_{n=1}^N$.
The motion of a topological soliton along a nanotube is conveniently described as the motion
of this one-dimensional anti-kink.

%----------------------------------------------------------------
\begin{figure}[t]
\begin{center}
\includegraphics[angle=0, width=1.0\linewidth]{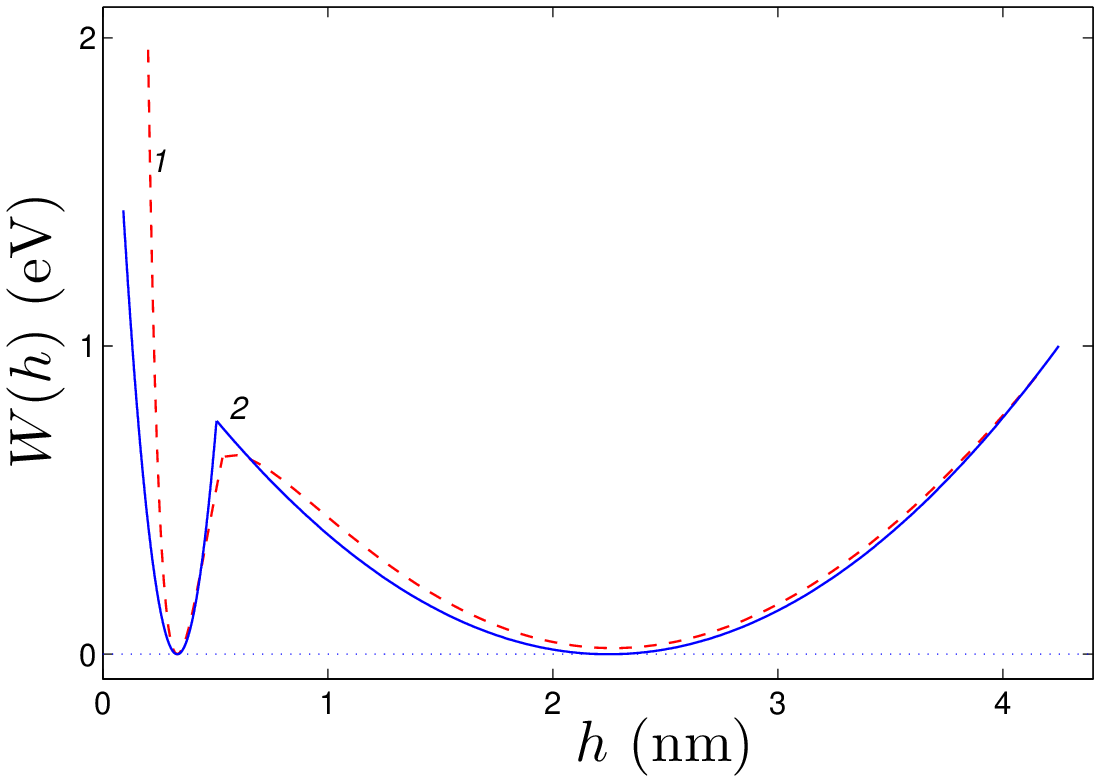} 
\end{center}
\caption{\label{fig6}\protect
View of the bistable potential with index $m=31$  $E(h)$ obtained from the energy form in the MD simulations
(curve 1) and its approximation by a bi-parabolic potential $W(h)$ (curve 2) with parameters
$h_1,h_0,h_2=0.3316$, 0.5058, 2.2473~nm, $K_1,K_2=50$, 0.5~eV/nm$^2$.
}
\end{figure}
%----------------------------------------------------------------
Simulation of nanotube dynamics shows that in the nanotube $(31,31)$ the topological soliton
moves to the left at $T<90$K and to the right at $T>90$K -- see Fig.~\ref{fig4}~(a).
Here, the temperature at which the direction of motion of the topological soliton is switched
$T_0=90$K (the collapsed state of the nanotube is the main one at $T<T_0$, and the open state
-- at $T>T_0$).
In the nanotube (32,32), the switching of the direction of motion of the soliton occurs already
at a higher temperature $T_0=600$K -- see Fig.~\ref{fig4}~(b). For wider nanotubes (33,33)
and (34,34), switching must occur at very high temperatures, so at $T\le 930$K, the topological
soliton always moves to the left, bringing the nanotube into a collapsed state.
The velocity of the soliton thus decreases with increasing temperature -- see Fig.~\ref{fig5}.

\section{Mechanism of the front speed temperature dependence}\label{MechTempDep}

The amplitudes of the deformations of the nanotube during the collapse from the hollow state
are sufficiently high. Description of such evolution using, for example, nonlinear Sanders-Koiter
thin shell theory should give us nonlinear equations which turn out to be too complicated
for analysis, as too many modes of deformations need to be taken into account \cite{Smirnov16}.
Therefore, we suggest an effective model representing the evolution of CNT to describe
the process of the collapse and a vice-versa process.

To describe the mechanism of the temperature dependence we consider the nanotube as coupled
unit-cells (rings of atoms) in a bistable potential defined by the properties of the nanotube
and by the interaction with the substrate. Let us represent the motion of each element as an
effective particle in a non-degenerate double-well potential -- see Fig.~\ref{fig3}. 
Than we can model the evolution of the CNT with a chain in a bistable on-site potential $E(h)$. 
The energy difference between the two potential minima can be defined as $\Delta E$.
For nanotube with chirality index (31,31) the two-well potential $E(h)$ can be approximated 
by a bi-parabolic potential
\begin{equation}
W(h)=
\begin{cases}
   \frac12 K_1 (h-h_1)^2~~\text{if $h\le h_0$,} \\
   \frac12 K_2 (h-h_2)^2~~\text{if $h> h_0$.}
 \end{cases}
\label{f12a}
\end{equation}
where potential minima are $h_1=3.316$~\AA, $h_2=22.473$~\AA, maximum point is $h_0=5.058$~\AA,
stiffness values are $K_1=0.5$~eV/\AA$^2$, $K_2=0.005$~eV/\AA$^2$ -- see Fig.~\ref{fig6}, $\Delta E=0$.

Half-openned (half-collapsed) state of the nanotube (31,31) may be described as kink (antikink)
in one-dimensional chain with double-well substrate potential (\ref{f12a}). Hamiltonian
of this chain
\begin{equation}
H=\sum_n\frac12\mu\dot{h}_n^2+\frac12\kappa(h_{n+1}-h_n)^2+W(h_n),
\end{equation} 
where effective mass of one chain cell $\mu\approx mM$, $m$ is chirality of the nanotube, $M$ is mass of carbon atom, (only about 1/4 part of the carbon atoms
in each transversal cell participate in the collapse of the nanotube).

\begin{figure}[t]
\begin{center}
\includegraphics[angle=0, width=1.0\linewidth]{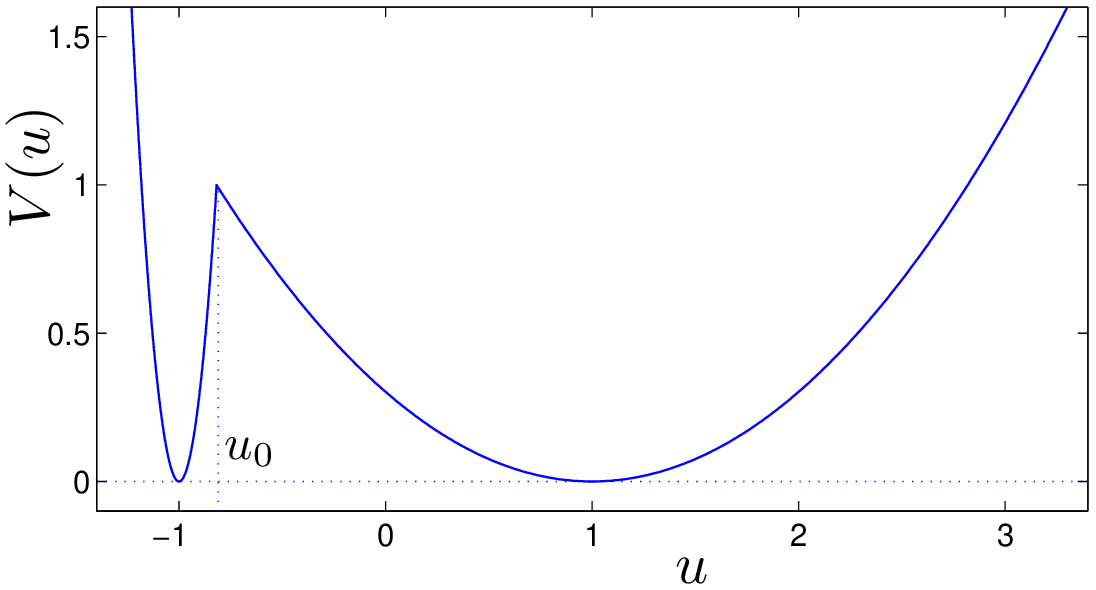} 
\end{center}
\caption{\label{fig6a}\protect
View of the bistable dimensionless potential V(u), $u_0=-0.8181$, $k_1=60.468$,
$k_2=0.60468$.
}
\end{figure}

For further analysis it is convenient to introduce the dimensionless displacements
$u_n=-1+(h_n-h_1)/h_d$, $h_d=(h_2-h_1)/2$ and dimensionless energy $E=H/E_{max}$, where $E_{max}=K_1(h_0-h_1)^2/2$.
Then dimensionless double-well potential $V(u)=W(h_1+(u+1)h_d)$ can be represented in the form:
\begin{equation}
V(u)=
\begin{cases}
   \frac12 k_1 (u+1)^2~~\text{if $u\le u_0$,} \\
   \frac12 k_2 (u-1)^2~~\text{if $u> u_0$,}
 \end{cases}
\label{f12b}
\end{equation}
where $u_0=-1+2(h_0-h_1)/(h_2-h_1)=-0.8181$, $k_i=(h_2-h_1)^2K_i/4E_0$, $i=1,2$, 
$k_1=60.468$, $k_2=0.60468$, see Fig.~\ref{fig6a}.

We suppose that the evolution along the CNT is smooth enough to apply the long-wave approximation:
%-------------------------------------- f9--------------------------------
\begin{eqnarray}
u_n = u,  \nonumber\\
u_{n+1} = u+a_0 u_x+\frac{a_0^2}{2}u_{xx},\\
u_{n-1} = u-a_0 u_x+\frac{a_0^2}{2}u_{xx},\nonumber
\label{f8a}
\end{eqnarray}
%-------------------------------------- f9--------------------------------
where $a_0=1$ is dimensionless distance between the two unit-cells, $x$ is a dimentionless longitudinal coordinate scaled by the longitudinal pitch $x_0=a$.
Under these assumptions we can model the dynamics of circumferential motion of the CNT
as an equation:
%-------------------------------------- f9--------------------------------
\begin{equation}
u_{tt}-c_0^2u_{xx}+\omega_0^2 \bar V'(u)=-\gamma u_{t}+\zeta(x,t),
\label{f9}
\end{equation}
%-------------------------------------- f9--------------------------------
where the $u$ is an effective coordinate having sense of the state coordinate,
which can be defined by the coordinate of center of mass of the moving part of nanotube, $c_0=\sqrt{\kappa a^2/\mu}$ is a longitudinal wave-speed. In general case the potential is non-degenerate, $u_1$ and $u_2$ are the local minima coordinates with the energy gap $\Delta\epsilon=|\bar V(u_1)-\bar V(u_2)|$. For nanotube with chirality (31,31) the potential in the dimensionless form $\bar V(u)$ can be taken from (\ref{f12b}) as $\bar V(u)=V(u)$, and $\omega_0=\sqrt{E_{max}/(\mu h_d^2)}$ is characteristic frequency of the motion, which yields:
\begin{equation}
\bar V(u)=
\begin{cases}
   \omega_{1}^2 (u-u_1)^2~~\text{if $u\le u_0$,} \\
   \omega_{2}^2 (u-u_2)^2~~\text{if $u> u_0$,}
 \end{cases}
\label{f13a}
\end{equation}
$u_1=-1$, $u_2=1$, $\omega_{1}=\sqrt{k_1/2}$ and $\omega_{2}=\sqrt{k_2/2}$ are local curvatures of the potential, which define frequencies of the motion in the vicinity of each minimum.  For nanotube with chirality (31,31) the characteristic frequency of the motion $\omega_0=1.148\cdot 10^{10}$ 1/s, $\omega_1=0.02273$, $\omega_2=0.00642$, see Fig.~\ref{fig6}. We neglect the energy gap between the two minima, for this case we take $\Delta\epsilon=0$. Let us note, that the more precise evaluation of the analytically obtained form of the potential $V(u)$ does not significantly change the results of the analysis. Therefore we keep simple bi-parabolic approximation.
%----------------------------------------------------------------
%----------------------------------------------------------------
%
The r.h.s of (\ref{f9}) defines the coupling to the equilibrium heat bath with a viscous
damping constant $\gamma$ and the Gaussian noise $\zeta(x,t)$ with zero mean and autocorrelation function:
$$
\langle\zeta(x,t)\zeta(x',t')\rangle =
2\gamma D\delta(x-x')\delta(t-t'),
$$
where $D=k_B T/A$, $A = h_d \mu$. In our calculations we used value $\gamma=\Gamma$ from the previous section.
The unperturbed kink for the degenerate potential $V(u)$ when $\Delta\epsilon=0$ in its implicit
form can be found as:
%-------------------------------------- f10--------------------------------
\begin{equation}
x-vt=\left(1-\frac{v^2}{c_0^2}\right)^{1/2} \frac{d}{\sqrt{2}}\int_{u_0}^{u_{1,2}(x-vt)}
\frac{du}{\sqrt{\bar V(u)}},
\label{f10}
\end{equation}
%-------------------------------------- f10--------------------------------
where $d=c_0/\omega_0$ is size of quasi-particle, constant speed is $v$, $|v|<c_0$,
$u_0$ -- maximum point of the potential [$\bar V(u_0)=\max\limits_{u_1<u<u_2} V(u)$],
the rest energy of the soliton is:
%-------------------------------------- f11--------------------------------
\begin{equation}
E_0=\omega_0 c_0 A \int_{u_1}^{u_2}\sqrt{2 V(u)} du,
\label{f11}
\end{equation}
%-------------------------------------- f11--------------------------------
rest mass is $M_0=E_0/c_0^2$. The anti-kink is defined from  (\ref{f10}) with reversed sign of
the right hand size.
The kink solution describes the transition of the nanotube profile from the collapsed state
to the open state, i. e. $u_{1,2}(x\rightarrow -\infty)=u_{1}$,
$u_{1,2}(x\rightarrow +\infty)=u_{2}$,
while the anti-kink solution corresponds to transition from the open state to the collapsed one,
$u_{2,1}(x\rightarrow -\infty)=u_{2}$, $u_{2,1}(x\rightarrow +\infty)=u_{1}$.
The initial conditions of the numerical simulations correspond to the anti-kink evolution.

The main factors defining the anti-kink velocity and its direction is the form of the two wells
of the potential and the energy gap $\Delta\epsilon$.
The phenomenon has an entropic nature and
is defined by the difference of profile of the both wells. To illustrate this phenomenon,
let us suppose that each element of our initial discrete chain is subject to the bistable
potential. The initial condition will correspond to the placement of the right side of the chain
to the left of the potential barrier, while the left side rests to the right of the barrier.
We can calculate the probability of the existence of the element in each of potential wells.
If the probability to find the element in the r.h.s. well is higher, than wave-front of the
switch between the two states will move to the right, and vice versa.
Following the approach described in the \cite{Costantini01} we define the wave-speed for
the almost degenerate case, when there energy gap is negligibly small. let us denote characteristic frequencies of the motion in the left and right valleys of the potential $\Omega_1=\omega_1 \omega_0$, $\Omega_2=\omega_2 \omega_0$.
In the frequency range   $(\Omega_2,\Omega_1)$  phonons yield net pressure of the antikink
to the right, while the phonons from the left side of the antikink are reflected back.
The result can be characterized by effective thermal force, which plays its role when
the local curvatures of potential close to both minima are different.
Using the approach \cite{Currie80}, we can characterize the motion of the kink or anti-kink
as a particle which undergoes Brownian motion with additional thermal force term in the Langevin equation:
%-------------------------------------- f12--------------------------------
\begin{equation}
M_0 \ddot{X}=-\gamma M_0 \dot{X}\mp(\Delta\epsilon+A_{th})+\xi(t),
\label{f12}
\end{equation}
%-------------------------------------- f12--------------------------------
where $X(t)$ is coordinate of center of mass of kink or antikink quasiparticle,
$M_0$ is its effective mass, $A_{th}$ represents effect of the thermal forcing,
and $\xi(t)$ is a Gaussian noise with zero mean and autocorrelation function
$\langle\xi(t)\xi(0)\rangle=2M_0k_BT/x_0 \delta(t)$.

To provide calculations we use approximate double-well potential $V(u)$ 
(see Fig.(~\ref{fig6})) for the case of $m=31$. Supposing that the two minima of potential correspond to the same energy value, we obtain the relation for the speed
of the anti-kink solution:
%-------------------------------------- f14--------------------------------
\begin{equation}
v_{th}=c_0\frac{kT}{2 E_0}\frac{(\omega_1-\omega_2)\omega_0}{\gamma},
\label{f14}
\end{equation}
%-------------------------------------- f14--------------------------------
where the energy of the quasi-particle is defined in (\ref{f11}).
Let us note, that the result is independent on $c_0$, as the value $E_0$ is proportional to $c_0$.
Using this fact let us simplify the relation as follows:
\begin{equation}
v_{th}=\frac{kT}{2 I A \omega_0} \frac{(\Omega_1-\Omega_2)}{\gamma},
\label{f14a}
\end{equation}
where constant 
$I= E_0/ (A \omega_0 c_0)=\int_{u_1}^{u_2}\sqrt{2 \bar V(u)} du$ 
is defined by the potential profile, $\Omega_1$, $\Omega_2$ are two characteristic frequencies of the motion in the vicinity of the stationary states, $\gamma$ is the friction parameter, term $A = h_d \mu$ defines the scale of the system.    

Comparison of the analytical speed-temperature dependence with results obtained numerically from the simulation of the nanotube with chirality (31,31) is presented in Fig.~\ref{fig7}.
Details of the analysis are presented in Appendix. 
%We see a fairly good correspondence for sufficiently low temperatures. For the higher temperature values the nonlinear character of the speed dependence is not predicted by the analysis, as the assumptions made for formulation of the problem for the kink quasiparticle may get inappropriate for the higher temperatures range. 
We see a fairly good correspondence
of the speed dependence, however, the starting point is not correct. This is due to the difference between the energy values at the two minima of the potential, which was neglected in our calculations.
However, such a good accordance of the obtained
analytically speed with the numerical results allows us to conclude that the considered phenomenon
has entropic nature.
%----------------------------------------------------------------
%\begin{figure}[t]
%\begin{center}
%\includegraphics[angle=0, width=\linewidth, keepaspectratio]{fg06.eps}
%\end{center}
%\caption{\label{fig6}
%View of the bistable potential at index $m=31$  $W(h)$ obtained from the energy form in the MD simulations
%(curve 1) and its approximation by a bi-parabolic potential (curve 2) with parameters
%$h_1,h_0,h_2=0.316$, 0.527, 2.247~nm, $K_1,K_2=50$, 0.59~eV/nm$^2$.}
%\end{figure}
%----------------------------------------------------------------
%\end{figure}
%----------------------------------------------------------------
\begin{figure}[t]
\begin{center}
\includegraphics[angle=0, width=1\linewidth]{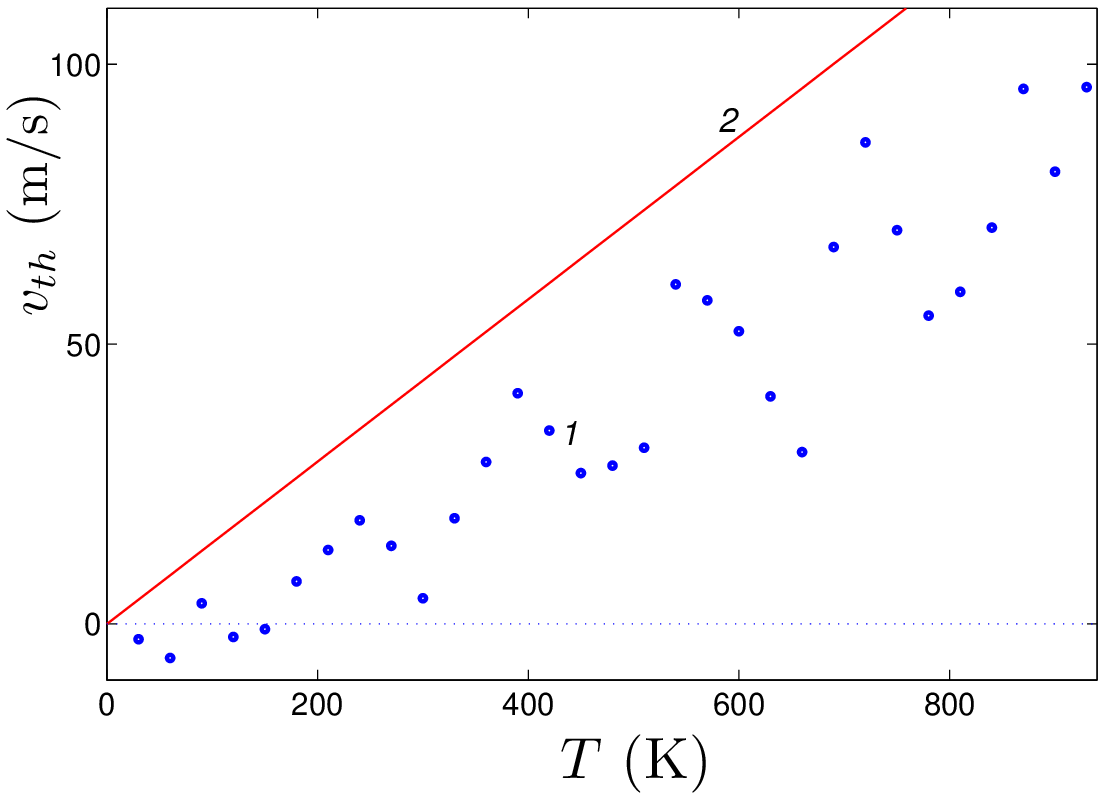}
\end{center}
\caption{\label{fig7}\protect
Speed of the soliton for temperature for the CNT index $m=31$ obtained in MD simulation (markers 1)
and comparison with the theoretical value form the degenerated bi-parabolic approximation (line 2).
}
\end{figure}
%----------------------------------------------------------------

\section{Conclusions}

In our work we consider the carbon nanotube on a surface with
interaction realized via effective Lennard-Jones potential. We
considered the evolution of the nanotube initially in the state with
one end collapsed and another end opened. The profile of the nanotube
demonstrates the soliton-like transition area from the opened to the
collapsed states. During the MD simulations we show the evolution of
the front as a soliton-like localized wave. We show that the direction
of the soliton (kink or anti-kink) depends on the radius of the
nanotube. We also demonstrate, that the temperature can affect
significantly the wave speed of the soliton. We find the energy
profile of the nanotube depending on the coordinate of the upper
middle-point of the nanotube for different index values of the
nanotube.

Using the effective model of the nanotube on the plane substrate connected with
thermostat we explain the dependence of the nanotube evolution as
effective chain of elements in a bistable potential. We prove the
entropic nature of the speed dependence on the temperature of the
system. The analytical results obtained in our asymptotic analysis
sufficiently well for such a sketch model correspond to those obtained via MD simulation.

\section*{Acknowledgements}
This work was supported by Program of Fundamental Researches of the Russian Academy of Sciences (project no. FFZE-2022-0009, state registration no.  122040500069-7).
Computational facilities for the work were provided by the Joint SuperComputer Center of the Russian Academy of Sciences.

%\newpage
\appendix
\section{Thermal speed of the soliton}
Here we present the detailed analysis of kink and anti-kink dynamics.
The small deformations of the soliton (\ref{f10}) can be studied in linear approximation
and expanded on the basis of eigenfunctions $\chi(x,t)=\psi(x) e^{-i\omega t}$:
\begin{equation}
-c_0^2 \psi_{xx}+\tilde V_0(x)\psi=\omega^2 \psi.
\label{f15}
\end{equation}
This equation can be represented as a one-dimensional Schrodinger equation with asymmetric
potential-well $\tilde V_0(x)=\omega_0^2 \bar V(u_{12}^{(0)}(x))$, where the $u_{1,2}^{(0)}$
is a static kink solution. The phonon modes with $\Omega_1^2>\omega^2>\Omega_2^2$ exert
an effective pressure on the kink to the left, while for the anti-kink the pressure acts to the right.
To estimate the effective thermal force $F_{th}$, acting to the kink or anti-kink let us consider
the diluted gas of kinks and anti-kinks with additional stopping tilt $F$ \cite{Costantini01}.
Using the transfer-matrix formalism \cite{Currie80} we obtain the free energy of the gas in thermal equilibrium as a corresponding eigenvalue problem:
\begin{eqnarray}
\hat H(u) \eta_n(u)=\epsilon_n(u),\nonumber \\
\hat H(u) =-\frac{\hbar^2}{2m_1}\frac{d^2}{du^2}+\bar V(u)+V_1-\frac{F}{\omega_0^2}u,
\label{f16}
\end{eqnarray}
where $m_1=(\hbar\omega_0c_0 A/k_BT)^2$ is an effective mass and $V_1$ is the energy offset depending on temperature. The difference between the lowest eigenvalues, i.e. the frequencies defining the curvature
in the two wells of the potential $\omega_{1,2}$ can be compensated by the stopping tilt:
\begin{eqnarray}
\frac{d_u F}{\omega_0^2}=\frac{\hbar}{2\sqrt{m_1}}\left(\sqrt{\bar V''(u_2)}
-\sqrt{\bar V''(u_1)}\right) \nonumber \\
=-k_BT \omega_0(\omega_1-\omega_2)/(2A c_0 \omega_0^2),
\label{f17}
\end{eqnarray}
where $d_u=u_2-u_1$ is distance between the two minima.
Consequently, the internal asymmetry of the double-well potential on the kink dynamics
can be estimated as the action of the external tilt $F_{th}$:
\begin{equation}
A_{th}=d_u F_{th}=\frac{k_BT\Delta\omega}{2Ad},
\label{f18}
\end{equation}
where $\Delta\omega=\omega_2-\omega_1$.
Using this fact, we suppose that the evolution of asymmetric kink or anti-kink can be described
by the Langevin equation as evolution of the Brownian particle.
\begin{equation}
M_0 \ddot{X}=-\gamma M_0 \dot{X}\mp(\Delta\epsilon+A_{th})+\xi(t),
\label{f19}
\end{equation}
where $X(t)$ is the coordinate of center of mass.
In the degenerate case energy difference between the two minima of potential $\Delta\epsilon=0$, and only the thermal effective force drives the kink or antikink
with the stationary speed to the left or to the right, consequently, with the speed:
\begin{equation}
\frac{v_{th}}{c_0}=\mp\frac{k_BT}{2 E_0}\frac{\Delta\omega \omega_0}{\gamma}
\label{f20}
\end{equation}
%bibliography{ApsNanotube6}% Produces the bibliography via BibTeX.
%bibliographystyle{ieeetr}

\end{document}